\documentclass{natureJLH}

\usepackage{graphicx}%
\usepackage{multirow}%
\usepackage{amsmath,amssymb,amsfonts}%
\usepackage{amsthm}%
\usepackage{mathrsfs}%
\usepackage[title]{appendix}%
\usepackage{xcolor}%
\usepackage{textcomp}%
\usepackage{manyfoot}%
\usepackage{booktabs}%
\usepackage{algorithm}%
\usepackage{algorithmicx}%
\usepackage{algpseudocode}%
\usepackage{listings}%
\usepackage{graphicx}
\usepackage[none]{hyphenat}
\usepackage{tikz}
\usepackage[normalem]{ulem} 
\usepackage{lineno}
\usepackage{amssymb}
\usepackage{amsmath}
\usepackage{caption}
\usepackage{booktabs,multirow}
\usepackage{multirow}
\usepackage{pdflscape}
\usepackage{xcolor}
\usepackage{hyperref}
\usepackage{pifont}
\usepackage{soul}
\usepackage{amsmath}
\usepackage{pdfpages}

\newcommand{\teff}{\ensuremath{T_{\mathrm {eff}}\,}}

\newcommand{\feh}{\ensuremath{[{\mathrm {Fe/H}}]\,}}

\newcommand{\mk}{\ensuremath{\mathrm M_K\,}}

\title{The formation and survival of the Milky Way's oldest stellar disk}

\author{Maosheng Xiang$^{\star1,2,3}$, Hans-Walter Rix$^{\star3}$,  Hang Yang$^{1,4}$,  Jifeng Liu$^{\star1,2,4}$, Yang Huang$^{1,4}$, and Neige Frankel$^{5,6}$ }

\begin{document}

\maketitle

\begin{affiliations}

\item Key Laboratory of Optical Astronomy, National Astronomical Observatories, Chinese Academy of Sciences, Beijing, 100101, China
\item Institute for Frontiers in Astronomy and Astrophysics, Beijing Normal University, Beijing, 102206, China
\item Max-Planck-Institut f\"ur Astronomie, K\"onigstuhl 17, D-69117, Heidelberg, Germany
\item School of Astronomy and Space Sciences, University of Chinese Academy of Sciences, Beijing, 100049, China
\item Canadian Institute for Theoretical Astrophysics, University of Toronto, 60 St. George Street, Toronto, Ontario M5S 3H8, Canada
\item David A. Dunlap Department of Astronomy and Astrophysics, University of Toronto, 
50 St. George Street, Toronto, Ontario M5S 3H4, Canada
\item Corresponding authors:  msxiang@nao.cas.cn; rix@mpia.de; jfliu@nao.cas.cn
\item Published in Nature Astronomy on 10 October, 2024  (https://doi.org/10.1038/s41550-024-02382-w); This manuscript is the accepted version. 
\end{affiliations}

\begin{abstract}

It remains a mystery when our Milky Way first formed a stellar disk component that survived and maintained its disk structure from subsequent galaxy mergers. We present a study of the age-dependent structure and star formation rate of the Milky Way's disk using high-$\alpha$ stars with significant orbital angular momentum that have precise age determinations. Our results show that the radial scale length is nearly independent of age, while the vertical scale height experienced dramatic evolution. A disk-like geometry presents even for populations older than $13$~Gyr, with the scale height-to-length ratio dropping below 0.5 for populations younger than 12.5~Gyr. We dub the oldest population that has maintained a disk geometry -- apparently formed over 13~Gyr ago -- {\em \textbf{PanGu}}. With an estimated present-day stellar mass of $\simeq2\times10^9$~$M_\odot$, {\em \textbf{PanGu}} is presumed to be a major stellar component of our Galaxy in the earliest epoch. The total present-day stellar mass of the whole high-$\alpha$ disk is $2\times 10^{10}M_\odot$, mostly formed during a distinct star formation rate peak of 11$M_\odot$/year around 11~Gyrs ago. A comparison with Milky Way analogs in the TNG50 simulations implies that our Galaxy has experienced an exceptionally quiescent dynamical history, even before the Gaia-Enceladus merger.

\end{abstract}

The formation of our Galaxy's disk has proceeded in two major phases that are almost disjoint in time \cite{Nissen2020, Xiang_Rix2022}. The earlier phase, which occurred in the first $\sim5$~Gyr, resulted in the formation of a rapidly enriched (or high-$\alpha$) old disk. Its stars are separated from those of the younger disk formed in the later phase in chemistry and kinematics \cite{Bensby2003, Haywood2013, Rix_Bovy2013, Hayden2015, Bland-Hawthorn2016}, while the detailed picture of transition and interplay between these disks is still under debate \cite{Bensby2007, Bovy2012, Kawata2016, Buck2020a, Sharma2021}.  

Recent studies with precise stellar age, kinematics and abundances suggest that the onset of the Galactic old stellar disk has been in the first billion years of the universe ($\tau\simeq$13~Gyr, or \textbf{$z\simeq7$}), when gases were in a relatively metal-poor (${\rm [Fe/H]}\lesssim-1$), chaotic state \cite{Xiang_Rix2022, Belokurov2022, Conroy2022}. The early formation of this old disk apparently preceded the epoch at which much of the stellar halo population was established via accretion of satellite galaxies \cite{Xiang_Rix2022}. Early galactic disks have also been revealed in distant universe at \textbf{$z>7$} by recent JWST observations \cite{Kartaltepe2023}. On the other hand, galaxy formation scenarios and simulations suggest that galaxy assembly in the earliest epoch is dynamically violent as merger events happen frequently \cite{Fakhouri2008, Sotillo-Ramos2022}, which is detrimental to the creation and maintenance of galactic stellar disks. It thus remains a mystery when our Milky Way first formed stars as a disk, and what the oldest disk component is that did not get destroyed by subsequent mergers.

While stellar chemistry and kinematics provide central information on the Milky Way's old disk formation, the most intuitive way to understand the disk's temporal evolution is to characterize how its structure depends on age of the stellar population \cite{Minchev2016, Mackereth2017, Xiang2018}. For this, precise stellar ages for a large sample of stars with a well-defined selection function are needed. Such stellar samples have become available only in recent years thanks to large surveys. Such attempts have been made using spectroscopic data sets from the SDSS/APOGEE survey \cite{Mackereth2017} or from the LAMOST survey \cite{Xiang2018}. However, due to relatively large uncertainties (20-40 per cent) in the spectroscopic age estimates, it is difficult for those studies to unravel details about the earliest formation history of the high-$\alpha$, old disk. Fortunately, stellar ages can now be estimated with much higher precision by combining the spectroscopic stellar parameters with precise astrometric data from the Gaia mission \cite{Prusti2016, Brown2018, Brown2021}. In particular, \cite{Xiang_Rix2022} derived ages, with a median precision of $\sim 8$ percent, for a quarter of a million subgiant stars using Gaia astrometric parallax and photometry, as well as spectroscopic parameters from the LAMOST Galactic surveys \cite{Zhao2012, Deng2012, Liu2014}. 

Here we present a reconstruction of the stellar \emph{mass} density profiles for different mono-age and mono-abundance sub-populations of the Galactic old, high-$\alpha$ disk, utilizing an updated version of the subgiant star sample of \cite{Xiang_Rix2022}. The spatial distribution of the updated subgiant sample in the $R$-$Z$ plane of the Galactic coordinate is presented in \textbf{Extended Figure~1}. The selection process of the high-$\alpha$, rotating disk stellar populations used in the current work is illustrated in \textbf{Extended Figure~2} and \textbf{Extended Figure~3}. 
We describe the stellar mass density distribution as $\rho(\vec{r} \vert \tau,\feh; \{\rho_0, H_R, H_Z\})$, where the model parameter $\rho_0$ refers to the disk mass density at the solar position, $H_R$ to the disk scale length and $H_Z$ to the scale height. These parameters are determined through forward modeling that rigorously incorporates the sample's selection function and does global parameter fitting with Markov-Chain Monte Carlo, with details given in the Methods section. An example of the model fits to the observed distance distributions for a few mono-age and mono-abundance sub-populations are shown in \textbf{Extended Figure~4}.    

\subsection{Structural parameters of the Galactic high-$\alpha$ disk}
\begin{figure}
\centering
\includegraphics[width=1.0\textwidth]{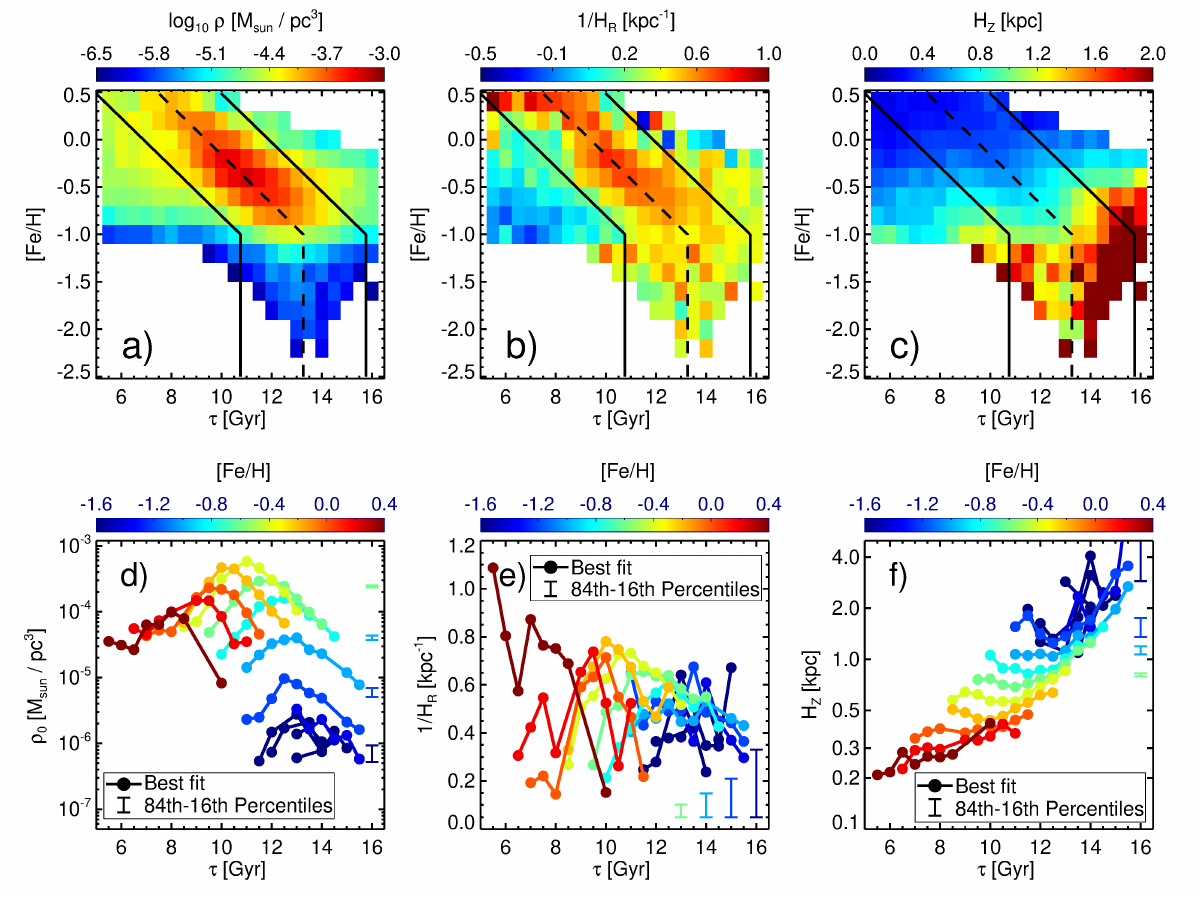}
\caption
{\textbf{Structural parameters for mono-age and mono-\feh populations of the high-$\alpha$ disk.} The upper panels show the distribution of best-fit parameters in the age-[Fe/H] plane. The dashed lines delineate parameter window in age-[Fe/H] where the high-$\alpha$ disk is expected to dominate. We only retain these subsamples for the subsequent analysis, as contamination by the low-$\alpha$ stellar populations (due to measurement errors) may dominate beyond. The lower panels show the best-fit parameters as a function of age for stellar populations in the selected age windows of the upper panels. The error bars shown in the figure represent the typical range of parameter estimates (i.e., 84th - 16th percentile of the integrated likelihood function) for several populations of different metallicity.} 
\label{fig:fig1}
\end{figure}

The best-fit parameters for each mono-age and mono-abundance populations are shown in Fig.\,\ref{fig:fig1}. Panel $a$ of the figure reveals a high-$\alpha$ disk sequence with high local density along a tight age-[Fe/H] relation from ${\rm [Fe/H]}\simeq-1$ 13\,Gyr ago to ${\rm [Fe/H]}\simeq0.5$ 8\,Gyr ago. The metal-poor tail of this sequence, in the range of $-2.5\lesssim{\rm [Fe/H]}\lesssim-1.0$, exhibits dramatically lower local stellar density. A more intuitive presentation of the local mass density trend is plotted in Panel $d$, which shows a dramatic jump of $\rho_0$ at ${\rm [Fe/H]}\simeq-1$ 13\,Gyr ago, as $\rho_0$ increases by about an order of magnitude from ${\rm [Fe/H]}\lesssim-1.2$ to ${\rm [Fe/H]}=-1.0$. This is an independent verification of the previous finding through the [Fe/H]--[$\alpha$/Fe] morphology that the star formation efficiency experienced a sudden increase at the early epoch, for a ${\rm [Fe/H]}\simeq-1.3$ \cite{Conroy2022}.  
 
Panel $b$ and panel $e$ suggest that the scale length of this old, high-alpha sequence is 1--3\,kpc, which evolves only marginally in the full range of metallicity and age, except for the extremely metal-rich (${\rm [Fe/H]}\simeq0.4$, $\tau<10$~Gyr) populations. However, the scale height decreases by an order of magnitude, from 2\,kpc for ${\rm [Fe/H]}<-1$ to 0.2\,kpc for ${\rm [Fe/H]}\gtrsim0$ (panel $c$ and panel $f$). For a constant [Fe/H], the scale height changes only moderately with age. These results suggest the high-alpha disk has experienced a much more dramatic structure evolution in the vertical direction than in the radial direction. 

We note that the Figure also displays significant populations of relatively young ($\lesssim8$~Gyr), metal-poor ($\feh\lesssim-0.5$) stars that our model fitting results a negative scale height (Panel $b$). These populations are likely contamination of low-$\alpha$ disk as a consequence of imperfect population distinction due to measurement errors in the [$\alpha$/Fe]. Therefore, in this work we confined our discussion about the high-$\alpha$ disk for populations in an age window along the distinguished, tight age-[Fe/H] sequence, as marked by the solid lines in the upper panels of Figure~\ref{fig:fig1}. Also note that here we have ignored the disk flaring effect in the modelling. Similar results are found after taking the flaring effect into consideration (see \textbf{\textbf{Extended Figure~5}}, and more detailed discussions in the Method section and the Supplementary file).

\subsection{The age-dependent disk structure}
\begin{figure}
\centering
\includegraphics[width=1.0\textwidth]{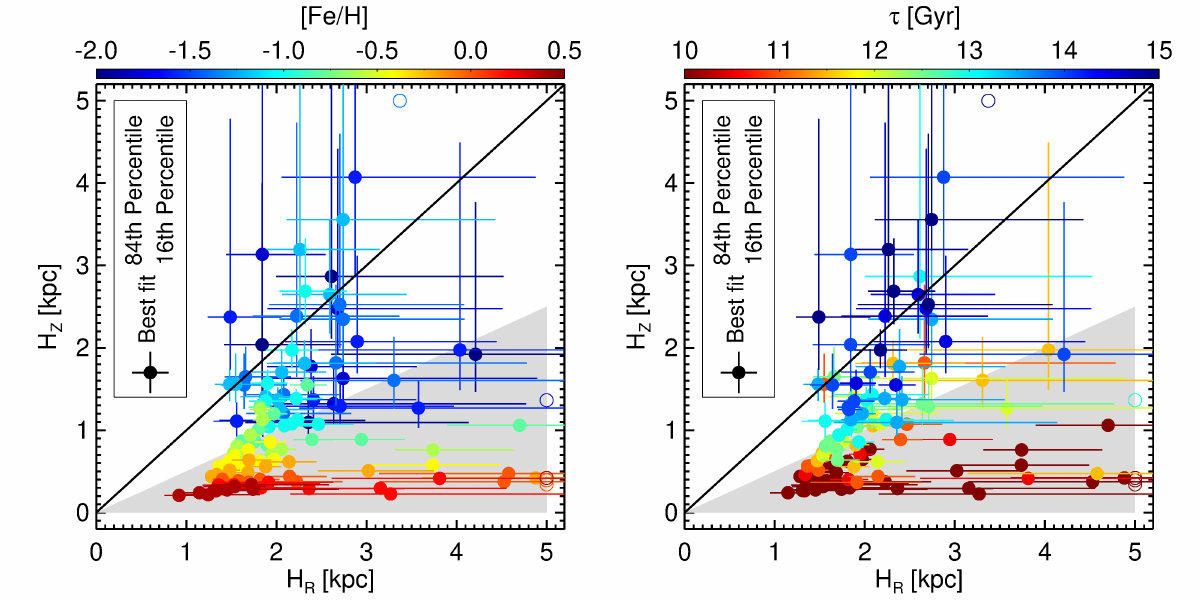}
\caption
{\textbf{Scale height versus scale length of the high-$\alpha$ disk.} The left and right panel are color-coded by metalicity and age of the stellar sub-populations, respectively. The open circles without error bars in the figure refer to results where the MCMC fit failed to yield a valid parameter constraint. The scale height is smaller than the scale length for nearly all components with $\tau\lesssim 13$~Gyr, and also for some sub-populations with even older ages. The shaded regions mark the parameter space where the scale height is smaller than half the scale length. Some of the most metal-poor ($\feh\lesssim-1.5)$ sub-populations exhibit smaller scale height than scale length, but many of them have ages younger than 13~Gyr. } 
\label{fig:fig2}
\end{figure}

Figure~\ref{fig:fig2} compares the scale length and scale height of the high-$\alpha$ disk for the mono-age and mono-abundance sub-populations. It illustrates that the scale height $H_Z$ correlates tightly with age. The vast majority of sub-populations of ages $\tau\lesssim13$~Gyr, or metallicity $\feh\gtrsim-1$, exhibit a disk-like geometry, with $H_Z$ smaller than $H_R$. Such a disk-like geometry even presents for older populations ($\tau\simeq13.5$~Gyr), while the majority of the oldest populations ($\tau\gtrsim14$~Gyr) have a scale height comparable to the scale length. The scale height drops below half the scale length for sub-populations with $\tau\lesssim12.5$~Gyr.

However, the metallicity dependence for populations of ${\rm [Fe/H]}\lesssim-1$ is less clear. Some of these metal-poor sub-populations exhibit scale heights comparable to the scale length, but a significant number of them, even for those of $\feh\lesssim-1.5$, can exhibit disk-like geometry. On the other hand, while the majority of the sub-populations with $\feh\simeq-1$ exhibit disk-like geometry, some of them may have a scale height comparable to the scale length. This complex, non-monotonic relation between the geometry and metallicity might indicate a complex early assembly and enrichment history of our Galaxy.

Interestingly, many of the most metal-poor sub-populations ($[{\rm Fe/H]}\lesssim-1.5$) that exhibit disk-like geometry tend to have younger ages ($\tau<13$~Gyr) than the populations of similar metallicity but with $H_Z\simeq H_R$. This implies that these metal-poor but younger disk-like populations may have experienced a different origin to the older populations of similar metallicity. It is possible that they are a part of accreted populations with co-planer orbits.

\subsection{Comparison with Milky Way analogs in TNG-50 simulation }

\begin{figure}
\centering
\includegraphics[width=0.8\textwidth]{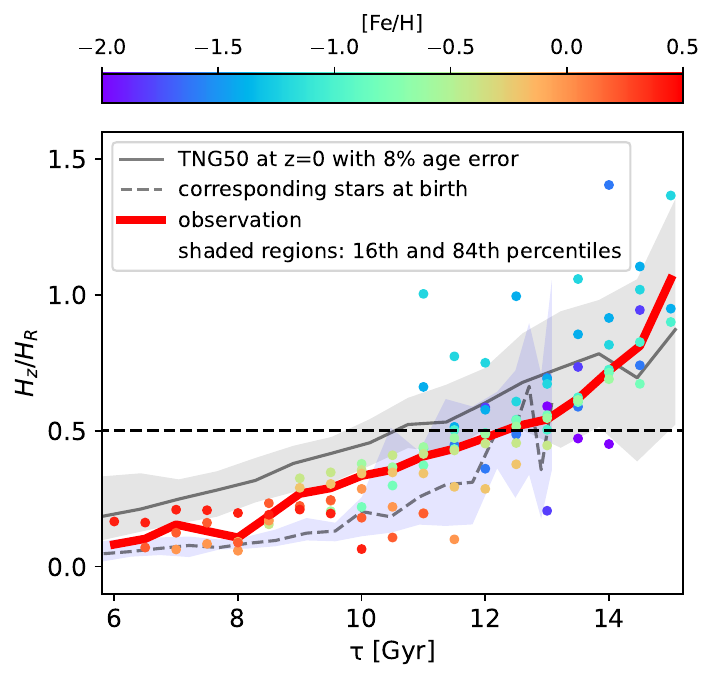}
\caption
{\textbf{Scale height-to-length ratio as a function of age, comparing Milky Way observations to TNG~50 simulations.} The dots are our measurements, color-coded by [Fe/H] of the stellar populations. Only populations with relative uncertainties in H$_z$/H$_R$  $<50$\% are shown. The solid red line is the mean observed  H$_z$/H$_R(\tau)$-relation marginalized over abundances. The solid and dashed lines in black show the average height-to-length ratio for Milky Way analogs in the TNG50 simulation, with the shaded regions indicating the variance among individual galaxies. The solid black line shows the stellar distribution morphology at present, after adding an age error of 8\% to the simulation output to mimic the observation. The dashed line shows the morphology \emph{at birth} of the corresponding population. Populations with formal ages older than the universe result from the age uncertainties.} 
\label{fig:fig3}
\end{figure}

In Figure~\ref{fig:fig3} we further analyze the age dependence of the high-$\alpha$ disk structure, looking at the scale height-to-length ratio, $H_Z$/$H_R$ as a function of stellar age. We compare the observational results with analogous quantities for 31 Milky-Way-like galaxies from the TGN50-1 simulation \cite{2019Nelson, 2019Pillepich}. These simulated galaxies were selected from the sample of \cite{Pillepich2023}, which contains 198 analogs of M31 or the Milky Way. In each of them, we choose \emph{in situ} stars whose $L_z > 500$ kpc.km/s and galactic distances from 6 to 12 kpc. When calculating the present $H_Z$/$H_R$ in each age bins, we incorporate a normal age error for all simulated stars. Here we only select the 31 Milky Way analogues (31\% among a total of 99) whose present-day $H_Z$/$H_R$-age trend resembles our Milky Way, i.e. have a smoothly decreasing $H_Z$/$H_R$ with age (solid grey line). For these 31 simulations, we can compare the present-day structure to the same stars' structure at birth (blue dashed line). The Figure illustrates that the simulated stellar populations can be born disk-like at the beginning, within the first Gyr of the Universe, but they are considerably thicker now than at birth due to dynamical heating. These results imply that even the oldest stars in our observation, which have $H_Z/H_R\simeq1$, might have initially formed in a disk-like structure but were subsequently destroyed by frequent galaxy mergers at the early epochs. They might now form the high angular momentum tail of the ``poor old heart of the Milky Way'' \cite{Belokurov2022,Rix2022}.   

Nevertheless, there is some slight difference in the present-day height-to-length ratio between the simulation and our measurements. In the simulations most components older than $\sim$10.5\,Gyr have $H_Z/H_R\gtrsim0.5$, while in the Milky Way this is only the case for stars of $\tau >12.5$\,Gyr. The Milky Way's $H_Z/H_R$ is below $1\sigma$ border of the simulated galaxies, which means that only $\lesssim 1/6$ of the simulated Milky Way-like galaxies have surviving disk populations with $H_Z/H_R$ comparable to the Milky Way. The reason for such a difference remains to be investigated. It may point towards a problem in the current simulations, but it may also be an indication that our Milky Way is extremely quiet so that the disk heating is not efficient compared to its analogs. We know that our Milky Way has experienced at least one prominent merger beyond the first 2~Gyrs, the  Gaia-Encelaus-Sausage \cite{Belokurov2018, Helmi2018, Helmi2020}, whose final phase was completed less than 12.5~Gyrs ago. This suggests that the impact of this early merger to our Milky Way's disk has been less traumatic than seen in the simulations.

\subsection{Star formation history of the high-$\alpha$ disk}
\begin{figure}
\centering
\includegraphics[width=0.8\textwidth]{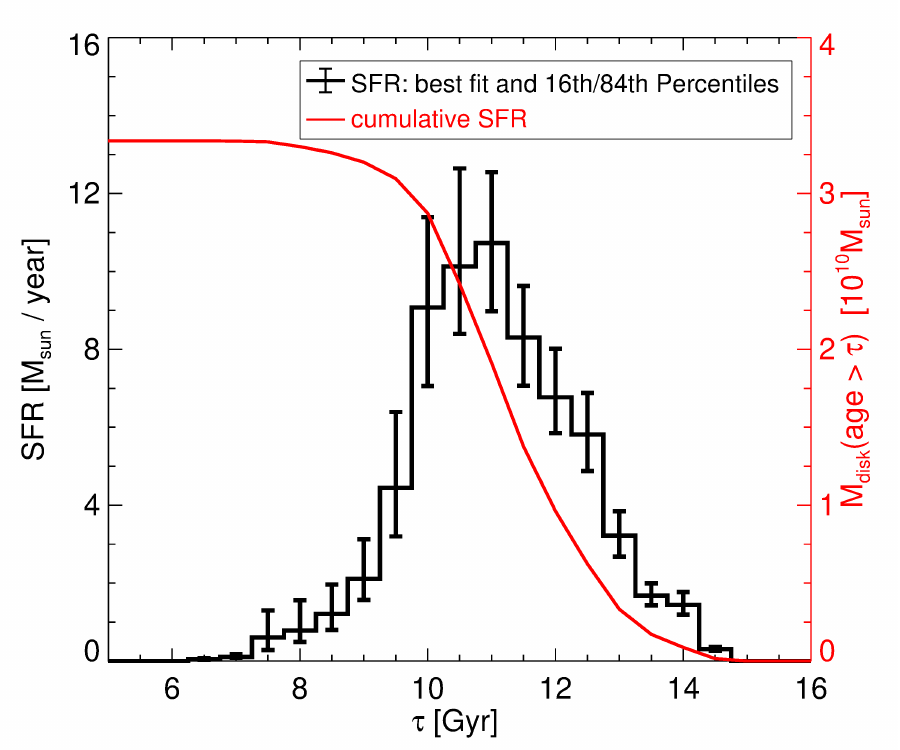}
\caption
{\textbf{Star formation rate and cumulative stellar mass of the high-$\alpha$ disk as a function of age.} The star formation rate is derived by integrating the structural parameters in the spatial range of $0<R<20$\,kpc, $-10<Z<10$\,kpc. At a given age, individual metallicity bins are co-added to generate the star formation rate shown in the figure. The error bars are estimated by propagating the fitting errors of the scale parameters. The cumulative stellar mass refers to the total stellar mass that the high-$\alpha$ disk formed before a given epoch. In fact, about 40\% of this mass has been lost \emph{via} stellar explosion, which turns the stars into gas and dust, providing the materials of chemical enrichment. Thus, one needs to multiply the number on the right-hand vertical axis by $\sim$0.6 to obtain the present-day disk stellar mass. }   
\label{fig:fig4}
\end{figure}

Our modeling also enables a rigorous estimate of the total stellar \emph{birth} mass of each mono-age and mono-abundance disk component by integrating the density profile over all radii and heights. This yields the star formation history of the high-$\alpha$ disk as a whole. Figure \ref{fig:fig4} shows this as the star formation rate. It started out at 2\,$M_\odot$ per year in the first few hundred Myr ($\tau>13$\,Gyr). This earliest high-$\alpha$ disk has built a total stellar mass of $3.7\times10^9M_\odot$, as of 13\,Gyr ago. Of this initial mass, $2.2\times10^9M_\odot$ remain as of now, after considering the mass loss by stellar explosion, which reduces 40\% of stellar mass according to the stellar evolution models. Most of these stellar masses, about $2.0\times10^9M_\odot$, were survived as disk that has $H_Z/H_R < 0.75$. The star formation rate increases to \textbf{6\,$M_\odot$} per year 12.5\,Gyr ago, and further to \textbf{11~$M_\odot$} per year 11\,Gyr ago. Such a peak star formation rate is consistent very well with both the chemical modeling \cite{Snaith2014} and the TNG50 hydrodynamic simulations, in which the Milky-Way-like galaxies exhibit a peak star formation rate of $\sim10$~$M_\odot$ per year at the early epoch \cite{Pillepich2023}.  
However, we find that the star formation rate then dropped quickly about 10\,Gyr ago, presumably because the gas was mostly depleted as a consequence of the high star formation rate before that epoch (but see \cite{Bonaca2020}). Ultimately, the high-$\alpha$ disk assembled a total of $2\times10^{10}M_\odot$ stellar mass (including remnants) that remains at present. 

\subsection{When did our Galaxy form its old disk?}

The onset of the Galactic disk has been suggested to happen at an early epoch with $\feh\simeq-1.3$, inferred either by modelling stellar distribution in the [Fe/H]-[Mg/Fe] plane \cite{Conroy2022}, or by characterizing the raise of stellar tangential velocities \cite{Belokurov2022}. Combining precise isochrone ages with our new geometry measurements shows that the (still surviving) Milky Way's high-$\alpha$ disk formation dates back to more than 13.5~Gyr ago. For many of the very oldest sub-populations in the sample ($\tau\simeq14$~Gyr), we do not find a clear disk-like morphology at present. It is, however, possible that these very oldest populations were once born as a disk but destroyed by early mergers, as also suggested by simulations. We therefore suggest that our Galaxy may have started as a disk. We dub the earliest disk formed in the first few hundred million years ($\tau>13$~Gyr) that has survived from early mergers and remains disk-like today as {\em \textbf{PanGu}}, the God who created heaven in Chinese mythology. Coincidentally, the two Chinese characters of the name are translated literally as old disk, as {\em \textbf{Pan}} means disk, {\em \textbf{Gu}} means old. The Galactic archeology results resonate with the substantive incidence of disk galaxies seen in $z>6$ JWST observations \cite{Kartaltepe2023}. 

Our results further suggest that {\em \textbf{PanGu}} has a present-day stellar mass of $M_*\simeq2\times10^9$~$M_\odot$. Even some more stellar mass may have born as disk along with {\em \textbf{PanGu}} but was destroyed by subsequent mergers. These destroyed disk populations may have partially contributed to the Galactic metal-poor (${\rm [M/H]}<-1.5$) heart \cite{Rix2022} (see also \cite{Belokurov2022} for {\em \textbf{Aurora}}). \cite{Rix2022} revealed a mass of $M_*\gtrsim10^8$~$M_\odot$ for the (metal-)\emph{Poor Old Heart} of our Galaxy. This is smaller than mass of the {\em \textbf{PanGu}}, implying that {\em \textbf{PanGu}} was likely a dominate component of our Galaxy at the earliest epoch. 

Our results also confirm that the vast majority of stars in the high-$\alpha$ disk formed at an extended burst that lasted from $\sim 12$ to $10$\,Gyr ago, which had been well established in previous work \cite{Bensby2007, Maoz2017, Silva_Aguirre2018, Xiang_Rix2022}. The present work now can add an actual rate estimate at the peak, of 11~$M_\odot$ per year, and also a decent estimate of the total stellar mass assembled in the high-$\alpha$ disk, of $2\times10^{10}M_\odot$ at present. This estimate of star formation rate, by direct age-dependent star counting, is remarkably consistent with chemical modeling \cite{Snaith2014} and hydrodynamical simulations \cite{Sotillo-Ramos2022, Pillepich2023}. 

\subsection{Mechanism for the growth of the old disk}
Our results have revealed that the disk scale height decreases by nearly an order of magnitude with decreasing age over the first 5~Gyr. At the same time, the variation of scale length is only moderate. This is likely to be a consequence of two effects: early on stars were born in a thicker disk (\emph{upside-down formation}; \cite{Bird2013, Bird21, Buck20, Lian2024}), or they may have gotten heated subsequently \cite{Bird2013, Bird21, Buck20, Ting19}. Both effects are seen in the simulations. 
Nevertheless, a final scenario for the assembly of the high-$\alpha$, old disk formation should explain,  besides the scale height-to-length ratio evolution, also other observations, for instance, the negative vertical metallicity gradients \cite{Hayden2014, Xiang2015, Wang2019}. Future work could carry out a comprehensive comparison between simulations or scenarios and observations.

As mentioned above, our results also hint that the assembly and chemical enrichment history in the first billion years are complex, rather than a simple, close-box gas enrichment that results in a single, monolithic age-metallicity relation. The most metal-poor stellar populations can exhibit disk-like geometry, which confirms, from a direct view, that the existence of a metal-weak (poor) thick disk \cite{Norris1985, Morrison1990, Beers2014, Sestito2020, Carter2021}. The results further show evidence that some of these metal-poor disk populations are probably results of early accretion events, as they tend to be younger than in-situ stellar populations of similar metallicity. 

Finally, a comparison of age evolution of the disk morphology with the TNG-50 simulations suggests only $\lesssim 1/6$ of the simulated Milky Way-like galaxies have surviving disk populations with $H_Z/H_R$ comparable to the Milky Way,  indicating our Milky Way has experienced a very dynamical quiescent history. The impact of early mergers, such as the Gaia-Encelaus-Sausage \cite{Belokurov2018, Helmi2018}, to our Milky Way's disk may have been less traumatic than seen in the simulations.

\clearpage

\section*{Methods}

\subsection{An updated LAMOST-Gaia subgiant star sample}
We build a sample of subgiant stars from the seventh data release (DR7) LAMOST Galactic survey (https://dr7.lamost.org) \cite{Yan2022} and third data release of the Gaia mission \cite{Brown2021}, following \cite{Xiang_Rix2022}. Subgiants are stars in a brief evolutionary phase where their core contracts and releases gravitational potential energy after the end of core hydrogen burning. Stars in this phase allow most precise age determination, as their luminosity is a direct and sensitive indicator of their ages \cite{Beers2022}. 

\cite{Xiang_Rix2022} had identified 247,104 subgiant stars from the $T_{\rm eff}$-$M_K$ diagram utilizing the LAMOST spectroscopic and Gaia DR3 astrometric data sets. These data afforded a median age uncertainty of only $\sim8$ per cent. Getting such a clean sample involved a number of steps: first, they discarded possible binary stars with a Gaia RUWE \cite{Lindegren2021, Belokurov2020} value larger than 1.2. They also further identified binary stars based on the excess of their parallax-based `geometric' $M_K$ w.r.t. their spectroscopic $M_K$, and discarded them from the sample. The `geometric' $M_K$ refers to $K$ band absolute magnitude inferred from distance modulus using Gaia parallax, and indicates the total luminosity of the binary system. Whereas the spectroscopic $M_K$ refers to that deduced from the normalized LAMOST spectra with a data-driven approach, which provides a good luminosity indicator of the primary star, particularly for equal-mass binary systems that the spectral lines of the individual components are indistinguishable \cite{Xiang2021, Xiang_Rix2022}. This method effectively eliminates binaries with high mass ratio ($\gtrsim0.7$), which may lead to erroneous age estimate if treating as single stars. This is different from, and thus a complementary to, the Gaia RUWE method, which may recognize binaries most effectively for a lower mass ratio. $M_K<0.5$ from the sample to eliminate potential contamination of core helium burning blue horizontal branch (BHB) stars. While these criteria leave a clean sample, they also result in incompleteness for stellar mass density reconstruction that needs to be characterized. 

To consider the effect of $\alpha$ element enhancement on the age estimation, \cite{Xiang_Rix2022} derived the stellar ages with the Yonsei-Yale (YY) isochrones \cite{Yi2001, Kim2002, Demarque2004} of different [$\alpha$/Fe], namely [$\alpha$/Fe]=0, 0.2, 0.4. To compute the best age, they combined the age estimates from these isochrones, weighted by Gaussian whose mean and $\sigma$ was taken from the LAMOST [$\alpha$/Fe] measurements. It was subsequently realized that this strategy is imperfect because the typical measurement error in the LAMOST [$\alpha$/Fe] ($<0.05$\,dex; \cite{Xiang2019}) is much smaller than the step of the isochrone grid (0.2~dex). 

Therefore, in the current work we update the Gaia and LAMOST subgiant star sample of \cite{Xiang_Rix2022}, by inheriting their overall approach,
but with several updates:
\begin{itemize}
   \item First, we include binary stars in our sample by discarding both the Gaia RUWE and $M_K$ criteria for binary elimination of \cite{Xiang_Rix2022}, but with a careful treatment of their age and distance estimates. For single stars, both geometric $M_K$ and spectroscopic $M_K$ are adopted for age estimation. For binary stars, as their geometric $M_K$ are over-luminous due to the contribution from the secondary, only the spectroscopic $M_K$ are used for a proper age estimation for the primary stars (thus the binary system). On the contrary, for distance estimates, it turns out that the geometric distance is a better choice for binaries, but the spectro-photometric distance can be wrong. So we only use the geometric distance for binaries. The inclusion of binaries add 70,228 stars to the \cite{Xiang_Rix2022} sample, among which 41,747 are recovered by removing the $M_K$ criterion, while the others are recovered by removing the Gaia RUWE criterion.  
   
  \item Second, we include subgiant stars with $M_K < 0.5$ in our sample but eliminate BHB star contamination by setting a metallicity cut of ${\rm [Fe/H]}>-0.8$. BHB stars are metal-poor halo populations and can be effectively eliminated by the metallicity cut. Meanwhile, subgiant stars within this magnitude range are expected to be young and metal-rich, so that they can be well survival from the metallicity cut. This update adds 6076 stars to the sample. However, we note that these updates do not make a significant impact to the present work as we only focus on the old disk populations ($\tau\gtrsim 8$~Gyr).
  
  \item Third, we compute three ages for each star using the YY isochrones of [$\alpha$/Fe]=0, 0.2, 0.4, respectively, and then deliver the final age estimate by linear interpolation to fit the measured [$\alpha$/Fe]. This treatment will slightly improve the age of stars with [$\alpha$/Fe]$\simeq$0.1 or [$\alpha$/Fe]$\simeq$0.3, the middle intervals of the isochrone grids, although on the whole the impact on the age estimates is minor.  
\end{itemize}
With these updates, we obtain a sample of 320,028 subgiant stars that have precision age, abundances and orbital parameters \cite{10.12149/101467}. The spatial distribution of these stars in the $R$-$Z$ plane can be found in \textbf{\textbf{Extended Figure~1}}.

\subsection{Selection of high-$\alpha$ stars}
We select high-$\alpha$ disk stars through their abundances and kinematics, using the same criteria as criteria as \cite{Xiang_Rix2022} (see also \textbf{\textbf{Extended Figure~2}}), 
\begin{equation}
  \begin{cases}
    \begin{cases}
    [\alpha/Fe] > 0.16, & \text{if $\feh < -0.5$}, \\
    [\alpha/Fe] > -0.16\feh + 0.08, & \text{if $\feh > -0.5$}, \\
    \end{cases} 
    \\ L_Z > 500~{\mathrm {kpc.km/s}}.
  \end{cases}
\end{equation}
These criteria lead to 86,001 high-$\alpha$ sample stars. In the subsequent analysis, we further discard stars younger than 5\,Gyr, as they may be dominated by low-$\alpha$ young disk stars that are misclassified due to uncertainties in abundance determinations. Also, a smaller fraction of them might be blue stragglers of the high-$\alpha$ disk population whose ages are incorrectly inferred by assuming single-star evolution isochrones \cite{Zhang2021}. 

It is found that the high-$\alpha$ stellar populations for ${\rm [Fe/H]}>-1.2$ is dominated by a rotating component with a peak at $L_Z\simeq1200$~kpc.km/s, while disk stars that were splashed into halo orbits by ancient mergers such as the Gaia-Sausage-Enceladus \cite{Bonaca2017, Bonaca2020, Belokurov2022} only contributed a small low-angular momentum ($L_Z\lesssim500$~kpc.km/s) tail  (\textbf{\textbf{Extended Figure~3}}). For ${\rm [Fe/H]}\lesssim-1.2$, the angular momentum shows a dominate peak at $L_Z\simeq-200$~kpc.km/s, which is mainly contributed by the accreted Gaia-Sausage-Enceladus stars. These accreted stars survived the [$\alpha$/Fe] selection because they have comparable [$\alpha$/Fe] values to the in-situ stars at such a low metallicity. Nevertheless, \textbf{\textbf{Extended Figure~3}} clearly shows the existence of a rotating stellar population for populations of $-1.4<{\rm [Fe/H]}<-1.2$, and $-1.6<{\rm [Fe/H]}<-1.4$. Such a low-metallicity, rotating stellar population has been revealed extensively in previous studies, and has been explained as the metal-weak tail of the thick disk populations \cite{Norris1985, Morrison1990, Beers2002, Beers2014, Carollo_etal2019, Sestito2019, Sestito2020, Carter2021, Yan2022MWTD, Bellazzini2024}. For the more metal-poor case of $-1.8<{\rm [Fe/H]}<-1.6$, it is harder to identify such a rotating population as the low-angular momentum population becomes too dominate. In this case, a significant portion of stars selected by the kinematic criterion $L_Z>500$~kpc.km/s might be just a part of the large-radius tail of the ``poor old heart" of the Milky Way \cite{Rix2022}, or the {\em \textbf{Aurora}} \cite{Belokurov2022}.

\subsection{Modelling the density distribution of subgiant stars} \label{sec:method} 
We now consider different subsets of subgiant stars, each in a narrow age and abundance range, mono-age and mono-abundance populations. For each of these, we reconstruct their mass density distribution {\em at birth}, $\rho(\vec{r} \vert \tau,\feh; \vec{\theta})$, characterized with a set of structural parameters $\vec{\theta}$. We note that stars will never stay at their birth place. When specifying stellar mass density  {\em at birth}, we need to distinguish from the stellar mass density {\em at present}, as the former reflects the underlying star formation history while the latter does not, because of mass loss due to stellar explosion. 

The connection between the underlying stellar \emph{mass} density and the \emph{number} density found in the sub-giant sample is complex and needs careful addressing: 
\begin{align}
\begin{split}
n_{\rm sg}(\vec{r} \vert \tau,\feh; \vec{\theta}) & = \frac{1}{\bar{m}_*(\vec{r}\vert \tau,\feh)} \int \rho(\vec{r} \vert \tau,\feh; \vec{\theta}) \\
& \times \xi(m_*^\prime \vert \tau,\feh) \times I(m_* \vert m_*^\prime, \tau,\feh) \\
& \times S(\vec{r}, m_* \vert \tau,\feh){\rm d}m_*^\prime,
\end{split}
\end{align}
where $\vec{r}:=\{l,b,d\}$, $m_*$ and $m_*^\prime$ refer to the present and initial stellar mass, respectively. $\xi(m_*^\prime)$ is the stellar initial mass function (IMF), $I(m_* \vert m_*^\prime, \tau,\feh)$ the initial and present mass relation, which can be delivered from stellar evolution models, and $S(\vec{r}, m_*)$ the sample's selection function. 
$\bar{m}_*$ is the effective mean mass of the subgiant stars, which is derived as
\begin{align}
\begin{split}
& \bar{m}_*(\vec{r}\vert \tau,\feh) = \\
& \frac{\int m_* \xi(m_*^\prime\vert \tau,\feh) \times I(m_* \vert m_*^\prime, \tau,\feh) \times S(\vec{r},m_*\vert \tau,\feh){\rm d}m_*^\prime}{\int \xi(m_*^\prime\vert \tau,\feh) \times I(m_* \vert m_*^\prime, \tau,\feh) \times S(\vec{r},m_*\vert \tau,\feh){\rm d}m_*^\prime}.
\end{split}
\end{align}
This expression shows that the effective mean mass of the sample's subgiant stars, $\bar{m}_*(\vec{r}\vert \tau,\feh)$, depends on their distance or position $\vec{r}$. This is because the survey's apparent magnitude limit includes bright  subgiants (high mass) within a far larger volume than a faint subgiant (low mass). For subpopulations selected to be mono-age and mono-abundance this effect is greatly reduced, as the absolute magnitude $M_K$ of subgiant stars of given age and metallicity is near-constant. 

We parameterize the spatial structure of each sub-populations' stellar mass density distribution as $\rho(\vec{r} \vert \tau,\feh; \vec{\theta})$, presuming that they follow an exponential profile in both the radial and vertical directions, 
\begin{align}
\rho(R,Z) = \rho_{R_\odot,0}\exp\left[-\alpha\cdot(R-R_\odot) \right] \exp \left(-\frac{\vert Z-Z_0\vert}{H_{Z}}\right),
\end{align}
where $\alpha:=1/H_R$ is the inverse of the disk scale length. \cite{Bovy2016}, \cite{Mackereth2017} and \cite{Lian2022} showed that the mono-age or mono-abundance stellar populations may show a complex radial profile, with some of them showing a density profile increasing outward. Using $\alpha:=1/H_R$ avoids divergences, but requires that $\alpha$ be positive or negative. We adopt a $R_\odot$ of 8.178\,kpc from the \emph{Gravity} measurement \cite{GRAVITY_Collaboration2019}. For simplicity, we also fix the value of $Z_0$ to be 0. Therefore, the model contains three free parameters, $\vec{\theta} = \{\rho_{R_\odot,0}, H_{R}, H_{Z}\}$.  

To link the stellar mass density with the stellar number density, we adopt the Kroupa IMF \cite{Kroupa2001} that we presume to be universal for all ages and abundances.  Recent studies have presented evidence for varying IMF with age and abundance \cite{Li2023}, but not in the relevant mass range $0.7~M_\odot < M_* < 1.5M_\odot$. In our case, ignoring the variation of IMF might cause small uncertainties in the derived stellar mass density $\rho_{R_\odot,0}$ and star formation rate. However, the scale parameters \{$H_R$, $H_Z$\} are not affected by the shape of the IMF.   
 
As in \cite{Xiang_Rix2022}, we adopt the YY isochrones to the initial-to-present mass relation, as well as to convert the fundamental stellar parameters \{$m_*, \tau, \feh$\} to photometric and spectroscopic observables. We note that because of systematic differences exist among different isochrones, to use the same isochrones as that for the age determination is essential to ensure precise mass estimation for the current purpose.

\subsection{The sample selection function}
The sample's selection procedures are implemented in the survey's footprint ($l$, $b$) and photometric color-magnitude diagram $(c, m)$ as well as subsequent cut off in the spectroscopic HR diagram $(T_{\rm eff}, M_K)$. The selection function thus need be further modeled as  
\begin{align}
\begin{split}
 S(l,b,d,m_*\vert \tau,\feh) & = S_L(l, b) \times S_L(c,m \vert l,b) \\
 & \times S_{\rm sg}(T_{\rm eff}, M_K \vert \tau, \feh) \\ 
 & \times \vert J(c,m, T_{\rm eff}, M_K; d, m_*, \tau, \feh, A)\vert \\
 & \times F(A\vert l,b,d),    
\end{split}
\end{align}   
where $S_L(l,b)$ is the selection function accounting for the survey's footprint, which is defined as 
\begin{equation}
\begin{cases}
 S_L(l, b) = 1, {\rm if~ covered ~by ~the ~LAMOST ~and ~Gaia ~footprint,} \\
 S_L(l, b) = 0, {\rm otherwise}
 \end{cases}
\end{equation} 
$S_L(c, m \vert l, b)$ is the LAMOST survey's selection function in the color-magnitude diagram, and is defined as 
\begin{align}
S_L(c, m \vert l, b) = N_{\rm LAMOST}(c, m \vert l, b) / N_{\rm parent}(c, m \vert l, b).
\end{align} 
Here $N_{\rm LAMOST}$ refers to the number of stars that have been targeted by LAMOST, have $S/N>20$, and have spectroscopic stellar labels derived with the DD-Payne \cite{Xiang2019}. We adopt the Gaia eDR3 $(BP-RP, G)$ diagram as the parent photometric color-magnitude diagram. 
The $(BP-RP, G)$ diagram is split into cells of $0.2\times0.5$\,mag to build a grid of selection function for each sky field $(l,b)$ of $2^\circ\times2^\circ$. The selection function of each star is then interpolated from the grid.

$S_{\rm sg}(T_{\rm eff}, M_K \vert \tau, \feh)$ is the part of the selection function that accounts for our choice of `subgiant' stars, defined as a region in the (\teff-\mk) diagram; its value is unity for stars within our adopted \teff and \mk border, and is zero for stars outside the border. 
   
Finally, $\vert J(c,m, T_{\rm eff}, M_K; d, m_*, \tau, \feh, A)\vert$ is the Jacobian matrix that converts the fundamental stellar physical parameters to observables. It also consistes of several parts
\begin{align}
\begin{split}
& \vert J(c,m, T_{\rm eff}, M_K; d, m_*, \tau, \feh, A)\vert  = \\
& \vert J_s(T_{\rm eff}, M_K; m_*, \tau, \feh)J_p(c,m; d, m_*, \tau, \feh, A)\vert,
\end{split}
\end{align}
where $J_s(T_{\rm eff}, M_K; m_*, \tau, \feh)$ converts the fundamental stellar parameters to the spectroscopic parameters with the YY stellar evolution models, and $J_p(c,m; m_*, \tau, \feh, d, A)$ converts the fundamental stellar parameters, distance, and extinction to the color and magnitude using distance modulus, 
\begin{align}
\vec{m} - \vec{M} = 5\log d-5 + \vec{A},
\end{align}
where the vectors refer to the corresponding magnitudes and extinction in the {G, BP, RP} bands. 

Combining all these elements, we define an effective selection function,  
\begin{align}
\begin{split}
 S_{\rm eff}(l,b,d \vert \tau,\feh) & =  \frac{1}{\bar{m}_*(l,b,d\vert \tau,\feh)}  \\
 & \int \xi(m_*^\prime \vert \tau,\feh) \times I(m_* \vert m_*^\prime, \tau,\feh) \\
 & \times S(l,b,d, m_* \vert \tau,\feh){\rm d}m_*^\prime, 
\end{split}
\end{align}
which links the incidence of subgiant stars in our sample to the stellar mass density distribution at birth,  
\begin{align}
n_{\rm sg}(l,b,d \vert \tau,\feh; \vec{\theta}) = \rho(l, b, d \vert \tau,\feh; \vec{\theta}) S_{\rm eff}(l,b,d \vert \tau,\feh).
\end{align}
As
\begin{align}
\rho(l, b, d \vert \tau,\feh; \vec{\theta}) = \rho(R, Z \vert \tau,\feh; \vec{\theta})\vert{J(R,Z; l,b,d)}\vert,
\end{align}
where $\vert{J(R,Z; l,b,d)}\vert$ is the Jacobian function for coordinate transformation from $(l, b, d)$ to $(R,Z)$.

\subsection{The 3D extinction map}
To model the expected color and magnitude for any envisioned subgiant stars we need a 3D extinction map of the Milky Way, $F(A\vert l,b,d)$. For the present purpose we build such a map using the LAMOST stars. As introduced in \cite{Xiang_Rix2022}, we derive the distance of the LAMOST DR7 stars by combining the spectro-photometric distance with the geometric distance from the Gaia parallax. We estimate the intrinsic stellar colors of the individual LAMOST stars using their spectroscopic parameters, and then derive the reddening and extinction with color excesses (see detailed introduction in \cite{Xiang_Rix2022}. Given the high precision of the spectroscopic parameters, internal uncertainty in the reddening $E(B-V)$ estimate for individual stars is only $\sim0.01$~mag in typical.    

For each of the $2^\circ\times2^\circ$ field, we build an extinction-distance relation by fitting the LAMOST stars with a series of function forms
\begin{align}
E_{B-V}(d) = p_0 + \frac{p_1}{1+\exp(-\frac{\log d + p_2}{p_3})} + p_4*\log d, 
\end{align}
\begin{align}
E_{B-V}(d) = p_0 + \frac{p_1\log d}{1+\exp(-\frac{\log d + p_2}{p_3})} + p_4*\log d, 
\end{align}
\begin{align}
E_{B-V}(d) = p_0 + \frac{p_1}{1+\exp(-\frac{\log d + p_2}{p_3})},
\end{align}
where $p_i$ are coefficients determined by the fitting. For each field, we choose the set of function that best fits the LAMOST data as the adopted extinction-distance relation. The $E(B-V)$ values are converted to extinction by multiplying the total-to-selective extinction coefficients, as described by \cite{Xiang_Rix2022}. 

In Supplementary Figure~1, we show a comparison of our 3D extinction with the extinction map of Schlegel et al. \cite[SFD;][]{Schlegel1998}, and with the 3D extinction map of Green et al. \cite[Bayestar19;][]{Green2019}. We found remarkable good agreement with these previous maps, with a mean difference and dispersion of only $-0.002$ and 0.008~mag, respectively, in the $E(B-V)$ difference between ours and the SFD map for stars with $|b|>20^\circ$, and a mean difference and dispersion of 0.013 and 0.063~mag, respectively, in the $E(B-V)$ difference between ours and the Bayestar19 map for stars with $|b|<10^\circ$. The SFD extinction map is a 2D integrated map, the good agreement with our 3D map is likely a reflection of the fact that most of the extinction are contributed by gas and dust in a local, thin disk. The relatively large dispersion with the Bayestar19 values for stars at low Galactic latitudes might be partially contributed by spatial inhomogeneity of the $E(B-V)$ distribution in a sky area of $2^\circ\times2^\circ$, i.e., our 3D map may fail to accurately describe the spatially inhomogeneous extinction in these regions. However, as an uncertainty of 0.06~mag in $E(B-V)$ estimates may only induce $\sim10$ per cent uncertainty in the distance modelling, it will not make a serious impact on our conclusions.   

\subsection{Model optimization with MCMC}
The disk structural parameters $\vec{\theta}$ can be estimated by taking the modelling of $n_{sg}$ in Equation (11) as a Poisson point process \cite[e.g.][]{Zari2023}. The likelihood to be maximized for each mono-age and mono-abundance population ($\tau$,\feh) can be expressed as

 \begin{align}
\begin{split}
 \ln L(\vec{\theta}) & = \sum_{i}\ln \int\rho(l,b,d\vert \vec{\theta})d^2\cos(b)S_{\rm eff}(l,b,d)\delta({l-l_i, b-b_i, d-d_i}){\rm d}l {\rm d}b {\rm d}d \\ 
& - \int \rho(l,b,d \vert \vec{\theta})d^2\cos(b)S_{\rm eff}(l,b,d){\rm d}l {\rm d}b {\rm d}d,
 \end{split}
 \end{align} 
where $\delta$ refers to the $Dirac$ function. In the first term, only the density $\rho(l,b,d\vert \vec{\theta})$ depends on $\vec{\theta}$, while the summation (in logarithmic scale) of the other quantities are constant, thus the equation can be further simplified \cite{Bovy2012, Bovy2016}. Given this likelihood function, the model parameters $\vec{\theta}$ are optimized via Markov Chain Monte Carlo (MCMC) method in {\sc IDL} environment \cite{Zobitz2011} (see Supplementary Figure~2). Only populations with $N>20$ are considered here to ensure robust structure fits.  

 The set of parameters that yield the maximal likelihood are adopted as the best-fitting parameters, while the parameters at 16th and 84th per cent of the integrated likelihood distribution function are adopted as the lower and upper error the parameters, respectively. In order to have an intuitive knowledge on how good the fits are, \textbf{\textbf{Extended Figure~4}} shows the distribution of stellar distance modulus for three mono-age and mono-abundance sub-populations as examples. On the whole, the fits with exponential disk models are satisfactory.

\subsection{Validating the assumption of a disk profile}
While a single exponential disk profile as we adopted (Equation~4) has been shown to be a good description for relatively metal-rich ($\feh\gtrsim-1$), high-$\alpha$ stellar populations \cite{Bovy2012, Bovy2016}, it is less clear if such a profile provides an optimal description for populations in the low metallicity tail of our high-$\alpha$, rotating stellar sample. We have therefore implemented a test by fitting the stellar distribution with a two-axial power-law ellipsoid halo profile  
\begin{equation}
\rho(R,Z) = \rho_{R_\odot, 0}\left[\frac{R_\odot}{\sqrt{R^2+(Z/q)^2}}\right]^{n},
\end{equation}
where $q$ is the ellipticity of the halo, $n$ the power index for density drop from the Galactic center. A spheroidal halo has $q=1$, while $q<1$ refers an oblate halo, and a smaller $q$ means a profile that is closer to a disk. For the Milky Way stellar halo population, canonical values for $q$ and $n$ are 0.62 and 2.77, respectively \cite{Juric2008}. 

We obtained a much smaller $q$ value (0.1--0.5) but much larger $n$ value ($\simeq6$) for all the mono-age populations than the canonical value for the stellar halo (see Supplementary Figure~3). The oldest population (15.5~Gyr) has a $q$ value comparable to the canonical stellar halo, but the $n$ value ($\simeq4.3$) is still significantly larger. We have also directly compared the radial and vertical variations of stellar density for different profiles (see Supplementary Figures~4 and 5). We found that the observed distribution for all the stellar sub-populations in our sample can be reasonably well described by the exponential disk profile, but very different from the canonical stellar halo profile.      

It has also been suggested that significant disk flaring phenomenon presents even for the high-$\alpha$ mono-abundance populations \cite{Mackereth2017, Yu2021, Lian2022} or for the oldest stellar population \cite{Xiang2018}, which may be related to its formation history \cite{Minchev2015}. In the above analysis we have opted to omit the flare, as the inclusion of it will induce more complexity, considering that the $H_Z/H_R$ is no longer an intuitive description of the disk geometry. However, to validate the effectiveness of such a simplification, we have examined results for model fits which includes the flaring effect, 
\begin{align}
\rho(R,Z) = \rho_{R_\odot,0}\exp\left[-\alpha\cdot(R-R_\odot) \right] \exp \left[-\frac{\vert Z-Z_0\vert}{f(R)\times H_{Z}}\right],
\end{align}
\begin{align}
f(R) = \exp\left[\beta(R-R_\odot)\right],
\end{align}
where an exponential function $f(R)$ is adopted to describe the flaring effect, i.e., the increase of scale height as a function of $R$. The strength of the flaring effect is described by the parameter $\beta$.

 The best-fitting parameters from this flared disk model for mono-age and mono-abundance populations are presented in \textbf{\textbf{Extended Figure~5}}. The results for $\rho_{R_\odot,0}$ (left panel), $H_R$ (middle-left panel), and $H_Z$ (middle-right panel) are quite similar to that presented in Figure~\ref{fig:fig1}, with the exception that the value of $1/H_R$ now becomes larger, i.e., typical scale length $H_R$ decreased from $\simeq2$~kpc in Figure~\ref{fig:fig1} to $\simeq1.5$~kpc in the current figure. This is because the existence of flaring effect, which exhibits a strength varying from $\beta\simeq0.05$ for the young ($\tau\simeq8$~Gyr), metal-rich sub-populations to $\beta\simeq0.2$ for the old ($\tau\simeq13$~Gyr), metal-poor sub-populations. 

In the presence of flare, the $H_Z/H_R$ varies with Galactocentric radius. In Supplementary Figures~6, we have presented detailed comparisons between $H_Z$ and $H_R$ for three different radii. We found that the scale height is smaller than the scale length for most of the sub-populations at the inner part of the disk. Particularly, for $R\lesssim6$~kpc, even the oldest stellar populations of $\tau>13.5$~Gyr may exhibit a clear disk geometry. As we expect that most stellar mass of the old, high-$\alpha$ disk populations are confined in a relatively small Galactic radius ($R\lesssim6$~kpc),  we conclude that imposing the disk flare in the modelling does not impact the conclusions in the main text.

\clearpage
\begin{figure}
    \center
    \includegraphics[width=1\textwidth]{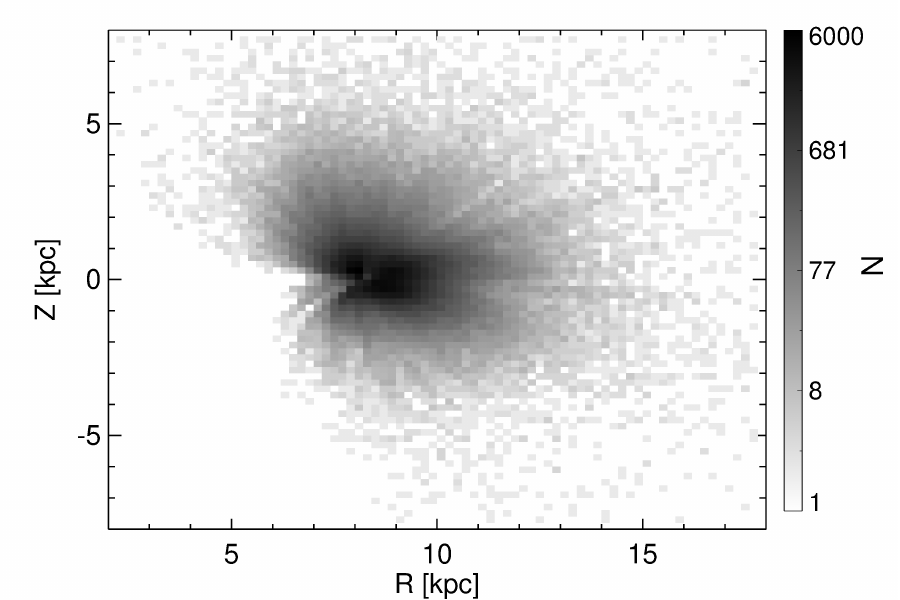}
    \caption*{{\bf Extended Data Fig. 1 $|$ Stellar number density distribution of the updated subgiant sample in the $R-Z$ plane.} The grey-scaled colors represent the stellar number in each 0.2~kpc $\times$ 0.2~kpc bin of the $R-Z$ plane.}
\end{figure}

\begin{figure}
    \center
    \includegraphics[width=1\textwidth]{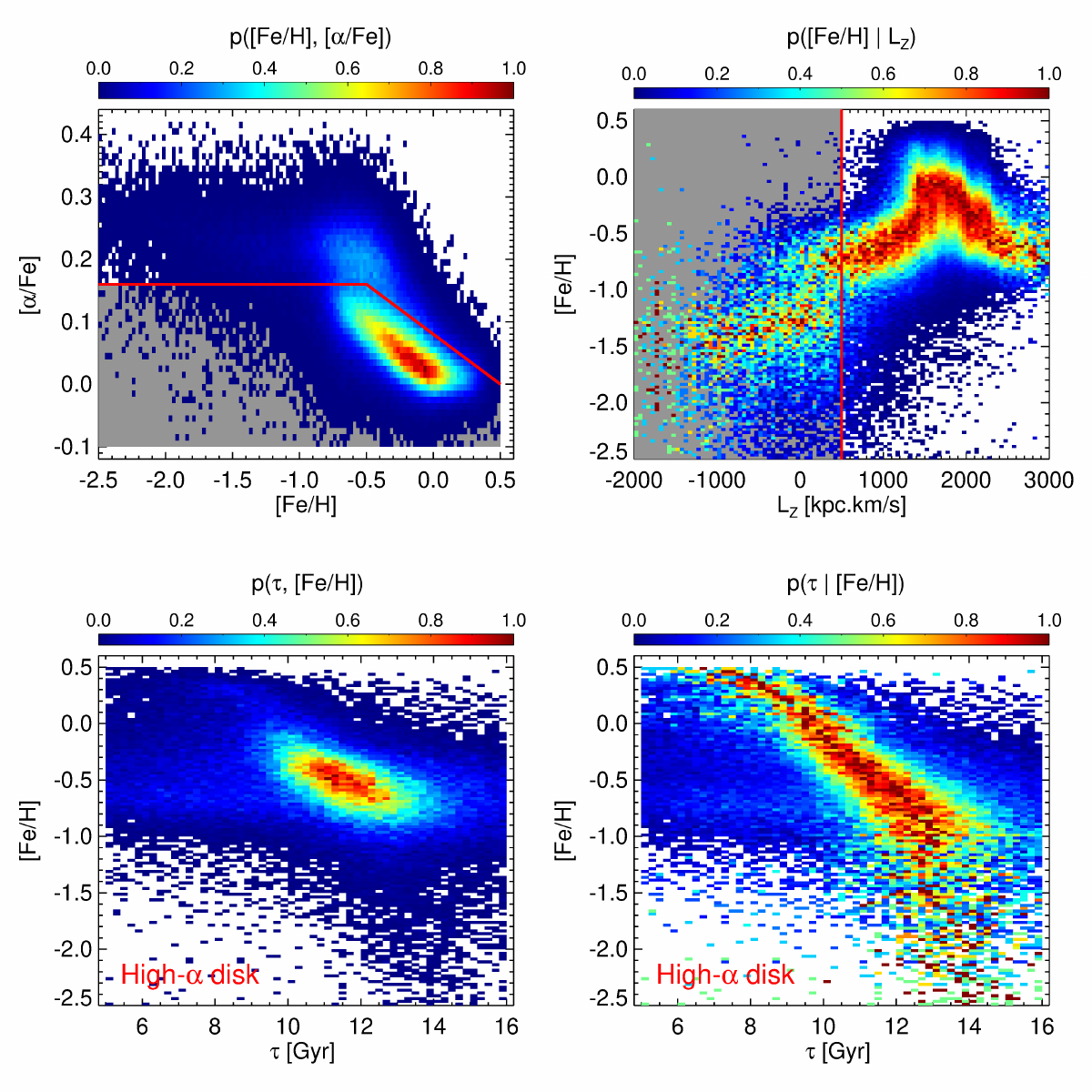}
    \caption*{{\bf Extended Data Fig. 2 $|$ Selection of the high-$\alpha$ disk star sample.} The upper panels show the selection of our high-$\alpha$ disk sample stars in the [Fe/H]-[$\alpha$/Fe] plane (upper left) and the $L_Z$-[Fe/H] plane (upper right), where $L_Z$ refers to the orbital angular momentum. To select high-$\alpha$ disk stars, both the low-$\alpha$ stars (shaded region in the upper left panel) and the kinematic halo stars (shaded region in the upper right panel) are discarded. The lower left panel shows the number density distribution of the resultant high-$\alpha$ disk star sample in the age-[Fe/H] plane, while the lower right panel is a row-normalized version of the lower left panel.}
\end{figure}

\begin{figure}
    \center
    \includegraphics[width=1\textwidth]{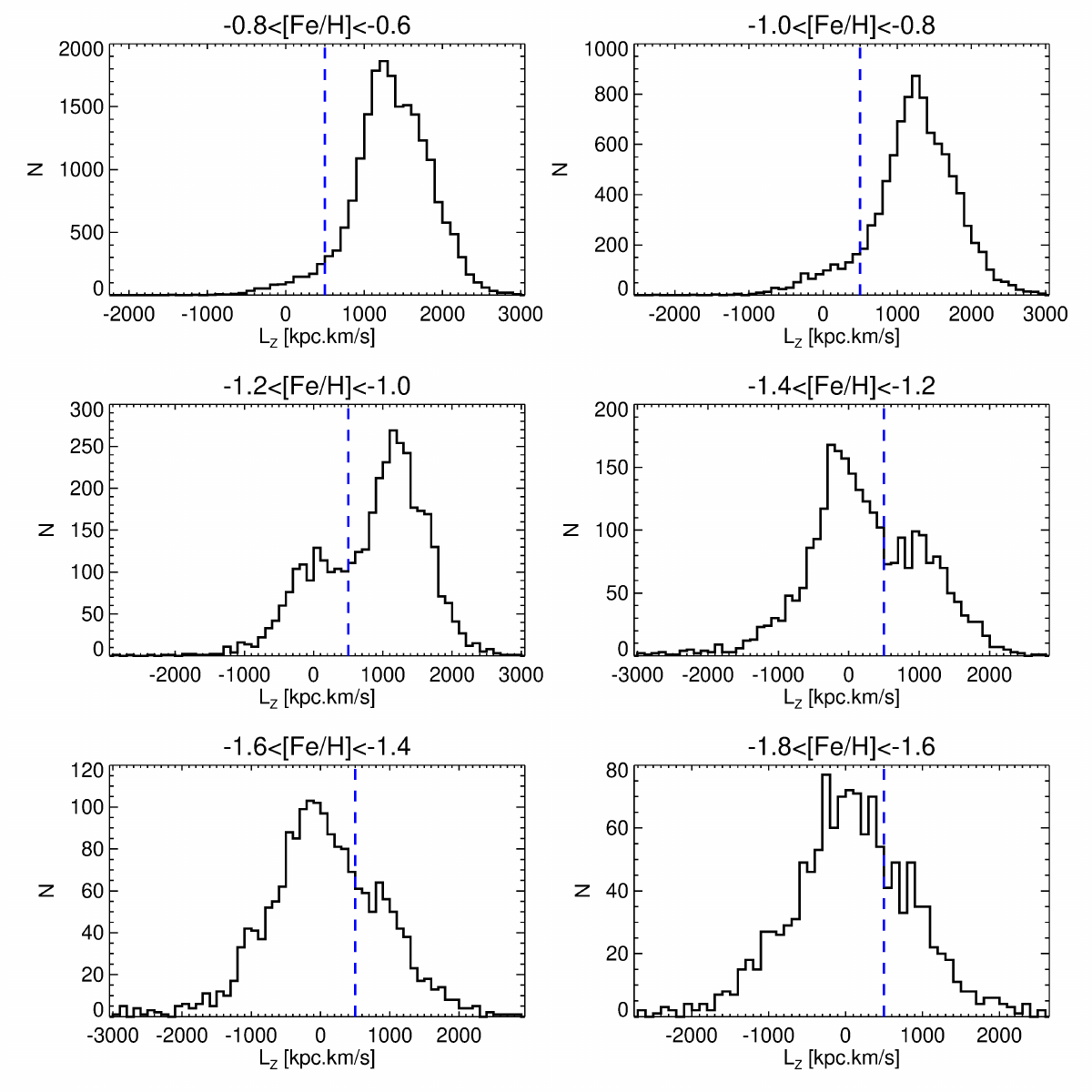}
    \caption*{{\bf Extended Data Fig. 3 $|$ Angular momentum distribution for high-$\alpha$ stellar populations.} Vertical dashed line delineates $L_Z = 500$~kpc.km/s, which we adopted for selecting the high-$\alpha$ disk sample stars.}
\end{figure}

\begin{figure}
    \center
    \includegraphics[width=1\textwidth]{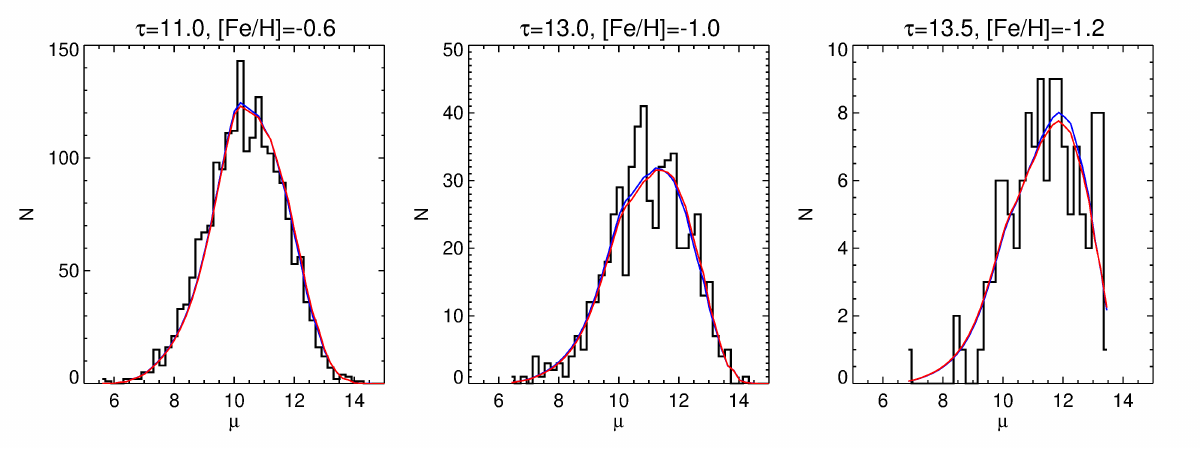}
    \caption*{{\bf Extended Data Fig. 4 $|$ Distribution of distance modulus ($\mu$) for three mono-age and mono-abundance sub-populations.} The red and blue curves are model fits to the observed distribution. The red curve represents the disk model with 3 parameters ($\rho_0$, $H_R$, $H_Z$), while the blue curve is the flared exponential disk model (see Method section).}
\end{figure}

\begin{figure}
    \center
    \includegraphics[width=1\textwidth]{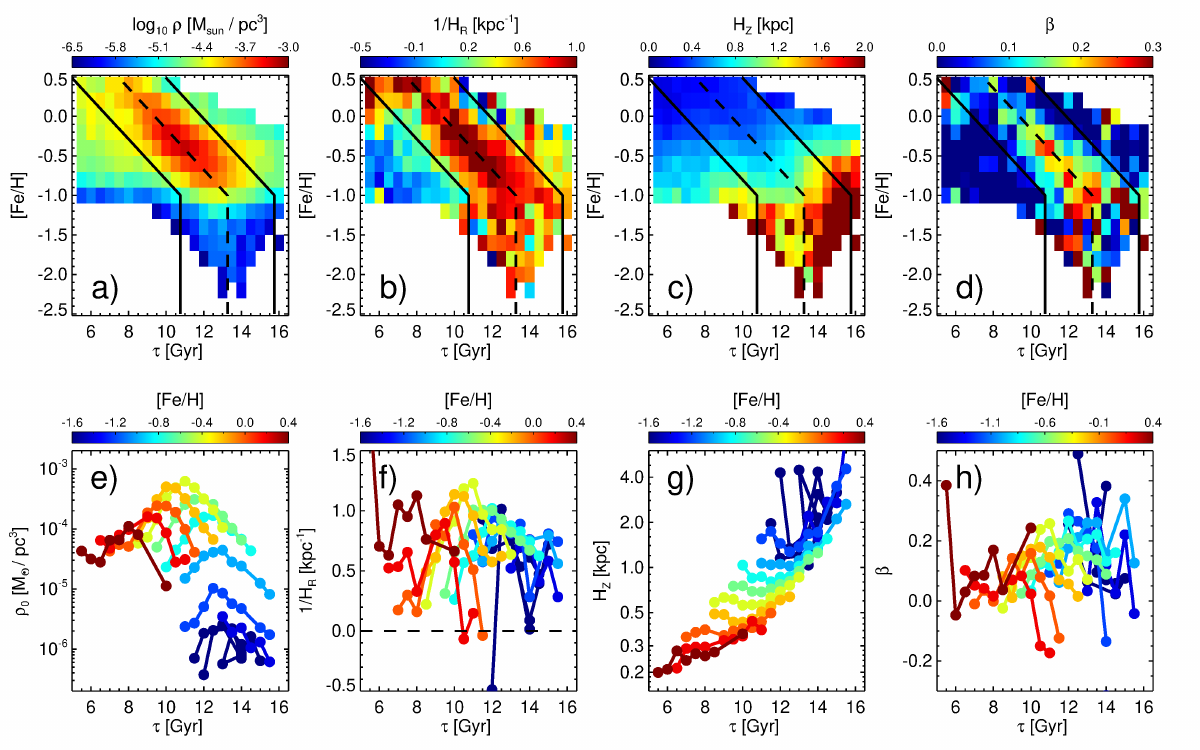}
    \caption*{{\bf Extended Data Fig. 5 | Structural parameters derived by fitting a flared disk model for mono-age and mono-[Fe/H] populations of the high-$\alpha$ disk.} The upper panels show the distribution of best-fit parameters in the age-[Fe/H] plane. From left to the right most panels are results for the local stellar density (a), the inverse scale length (b), the scale height (c), and the flare strength (d), respectively. The dashed lines delineate parameter window in age-[Fe/H] plane where the high-$\alpha$ disk is expected to dominate. We only retain these subsamples for the subsequent analysis, as contamination by the low-$\alpha$ stellar populations (due to measurement errors) may dominate beyond. The lower panels show the best-fit parameters as a function of age for stellar populations in the selected age windows of the upper panels, for the local stellar density (e), the inverse scale length (f), the scale height (g), and the flare strength (h).}
\end{figure}

\clearpage

\section{Data Availability}
The updated LAMOST and Gaia subgiant star catalog used in this work are publicly available via https://doi.org/10.12149/101466 \cite{10.12149/101467}. Additional data are available from the corresponding author.

\section{Acknowledgements}
MX acknowledges financial support from the National Key R\&D Program of China through grant No. 2022YFF0504200 and the NSFC through grant No.2022000083. HWR's research contribution is supported by the ERC Grant Agreement no.[321035]. JFL acknowledges support from the NSFC through grant Nos. of 11988101 and 11933004, and support from the New Cornerstone Science Foundation through the New Cornerstone Investigator Program and the XPLORER PRIZE.

This work has used data products from the Guoshoujing Telescope (LAMOST). LAMOST is a National Major Scientific Project built by the Chinese Academy of Sciences. Funding for the project has been provided by the National Development and Reform Commission. LAMOST is operated and managed by the National Astronomical Observatories, Chinese Academy of Sciences. The LAMOST website is https://www.lamost.org. 
 
This work has made use of data products from the European Space Agency (ESA) space mission Gaia. Gaia data are being processed by the Gaia Data Processing and Analysis Consortium (DPAC). Funding for the DPAC is provided by national institutions, in particular the institutions participating in the Gaia MultiLateral Agreement. The Gaia archive website is
https://archives.esac.esa.int/gaia. 

This work has made use the TNG50 simulation data. The TNG50 website is 

https://www.tng-project.org/data/milkyway+andromeda/.

\section{Author Contributions Statement}
M.~Xiang and H.-W.~Rix developed the initial idea of the project and the math of the forward modelling. M. Xiang conducted the data sample construction,  code implementation, and lead the data analysis with significant contribution from J. Liu and H.-W.~Rix. H. Yang and M. Xiang conducted the comparison of the observation results with the TNG50 simulation. M. Xiang and H.-W.~Rix wrote the manuscript, with revisions from N. Frankel, J. Liu, H. Yang, and Y. Huang.

\section{Competing Interests Statement}
The authors declare no competing interests.

\clearpage
\textbf{Supplementary file}

\textbf{1, Validating the 3D extinction map} 

\textbf{Supplementary Figure 1} shows a comparison of 3D extinction derived in the current work (see Method section of the online article) with literature. The left panel shows the $E(B-V)$ difference between our map and the (SFD; \cite{Schlegel1998}) extinction map for stars at high Galactic latitudes ($|b|>20^\circ$). It shows a very good consistency, with a mean difference of $-0.002$~mag and a dispersion of only 0.008~mag, validating the high precision of the 3D extinction map constructed in the current work. The SFD extinction map is a 2D map that has integrated interstellar extinction of the Milky Way along the line of sight. At high Galactic latitudes, it is a good approximation for extinction to the stars as most of the extinction are caused by gas and dust in the local, thin disk. However, the SFD map cannot accurately tell the extinction to a star at low Galactic latitudes as the interstellar extinction there is strongly dependent on the distance. Therefore in the right panel we compare our extinction to the 3D extinction map of (Bayestar19; \cite{Green2019}) for stars at low Galactic latitudes ($|b|<10^\circ$). It shows a mean difference and dispersion of 0.013 and 0.063~mag, respectively. An uncertainty of 0.06~mag in the $E(B-V)$ estimates may induce $\sim10$ per cent uncertainty in the distance modelling.   

\begin{figure}
    \center
    \includegraphics[width=1\textwidth]{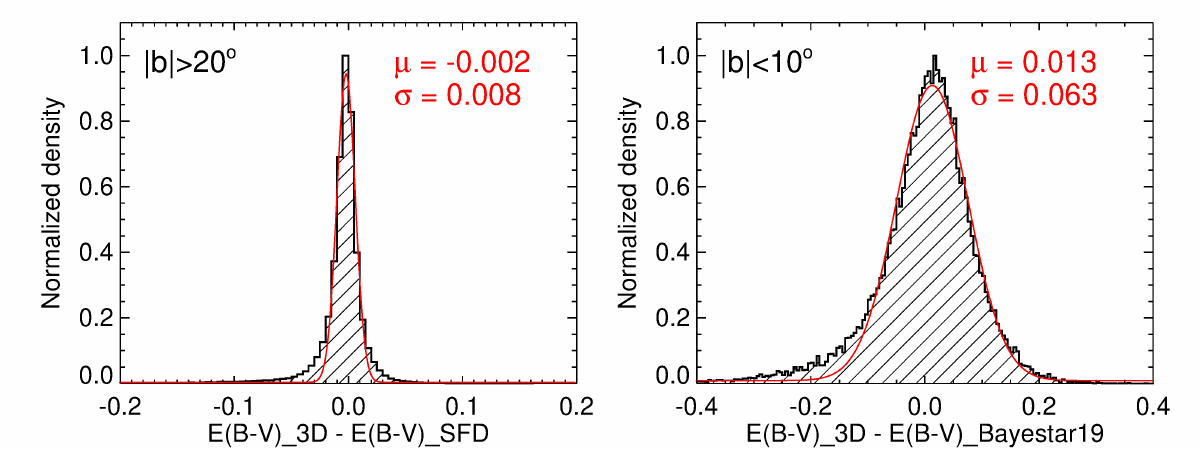}
    \caption*{{\bf Supplementary Figure 1 $|$ Distribution of $E(B-V)$ differences between our 3D map and literature.} The left panel the shows $E(B-V)$ differences with the (SFD; \cite{Schlegel1998}) map for stars at high Galactic latitudes. The right panel shows $E(B-V)$ differences with the (Bayestar19; \cite{Green2019}) 3D map for stars at low Galactic latitudes. Mean deviation and dispersion deduced from a Gaussian fit (red curve) to the distribution are marked.}
\end{figure}

\textbf{2, Validating the assumption of a disk profile}

We have adopted an exponential disk profile to fit the observed stellar distribution in the forward modelling. We adopt the Monte-Carlo Markov-Chain (MCMC) method for determination the parameter fitting. \textbf{Supplementary Figure 2} shows an example of the MCMC fitting.

\begin{figure}
    \center
    \includegraphics[width=0.8\textwidth]{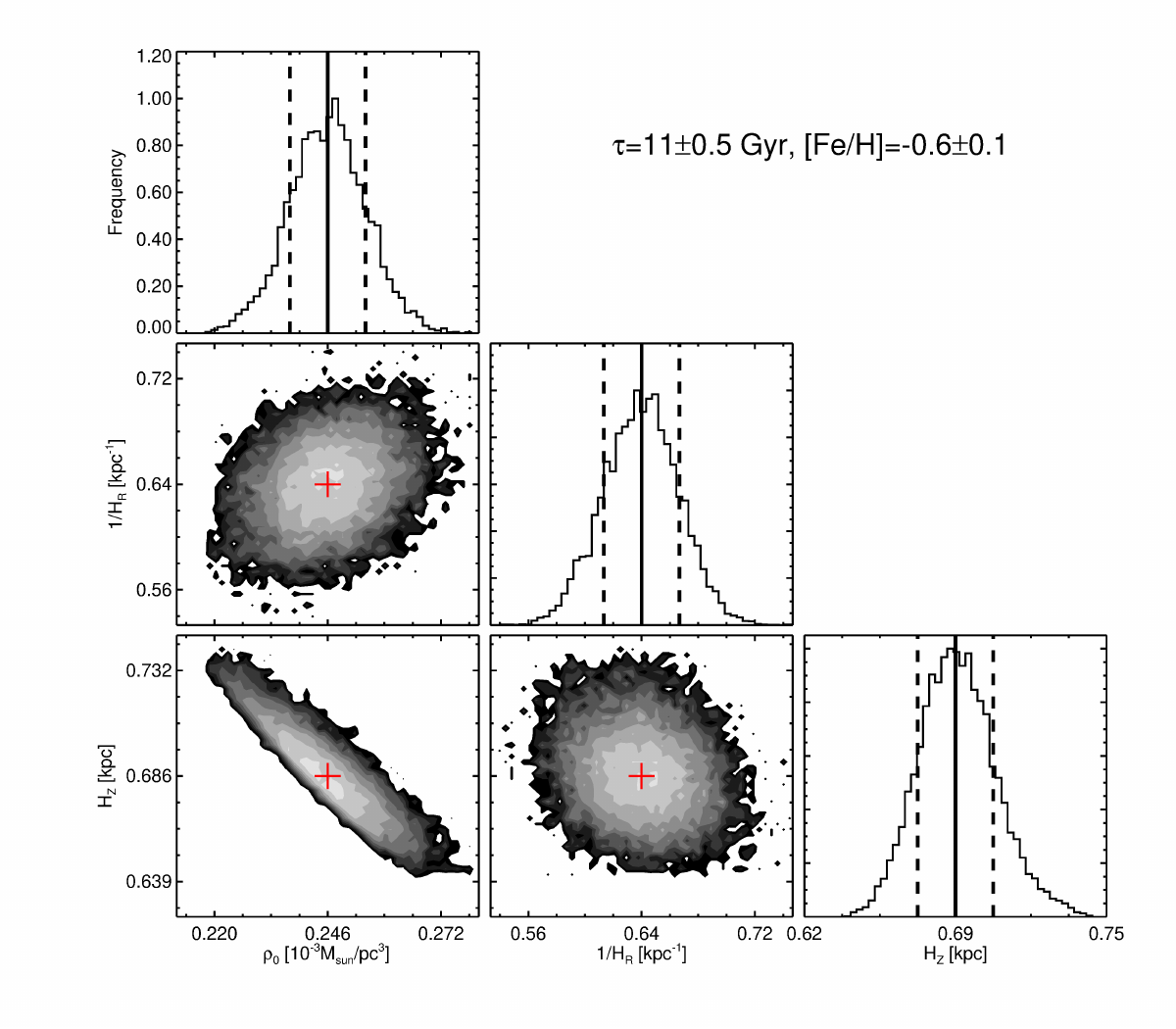}
    \caption*{{\bf Supplementary Figure 2 $|$ An example of MCMC fitting to the disk profile.} The results for stellar sub-population of $\tau=11$\,Gyr and $\feh=-0.6$ are shown. The vertical solid line and the red plus show the best-fitting parameters. The vertical dashed lines show the 16th and 84th percentiles of the posterior probability distribution.}
\end{figure}

Beyond this profile, we have also examined a two-axial power-law ellipsoid halo profile  
\begin{equation}
\rho(R,Z) = \rho_{R_\odot, 0}\left[\frac{R_\odot}{\sqrt{R^2+(Z/q)^2}}\right]^{n},
\end{equation}
where $q$ is the ellipticity of the halo, $n$ the power index for density drop from the Galactic center. A spheroidal halo has $q=1$, while $q<1$ refers an oblate halo, and a smaller $q$ means a profile that is closer to a disk. For the Milky Way stellar halo population, canonical values for $q$ and $n$ are 0.62 and 2.77, respectively \cite{Juric2008}. 

The resultant halo profiles for mono-age populations of our sample are shown in \textbf{Supplementary Figure 3}. Compared to the canonical values for the Milky Way stellar halo, we obtained a much smaller $q$ value (0.1--0.5) but much larger $n$ value ($\simeq6$) for all the mono-age populations. The oldest population (15.5~Gyr) has a $q$ value comparable to the canonical stellar halo, but the $n$ value ($\simeq4.3$) is still significantly larger. 
\textbf{Supplementary Figure 4} and \textbf{Supplementary Figure 5} directly compare the radial and vertical variations of stellar density for different profiles. It turns out that the observed distribution for all the stellar sub-populations can be reasonably well described by the exponential disk profile in the radial range covered by our data ($6\lesssim R \lesssim14$~kpc), but very different from the canonical stellar halo profile.      

\begin{figure}
    \center
    \includegraphics[width=1.0\textwidth]{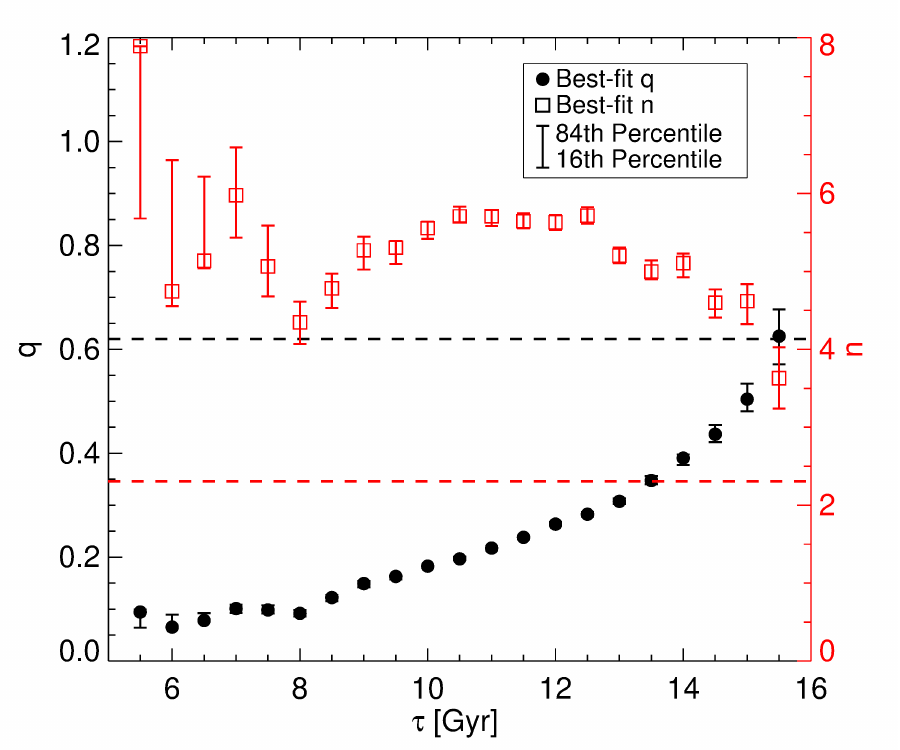}
    \caption*{{\bf Supplementary Figure 3 $|$ Ellipticity $q$ and power index $n$ derived by fitting the stellar distribution of mono-age populations with a two-axial power-law ellipsoid halo profile (Equation~17 in the article).} Dashed lines delineate the literature values for the canonical stellar halo, $q=0.62$ and $n=2.77$ \cite{Juric2008}. Fitting the distribution of mono-age populations in our sample with the halo profile model yields too large $n$ but too small $q$ compared to the canonical stellar halo.}
\end{figure}

\begin{figure}
    \center
    \includegraphics[width=1.0\textwidth]{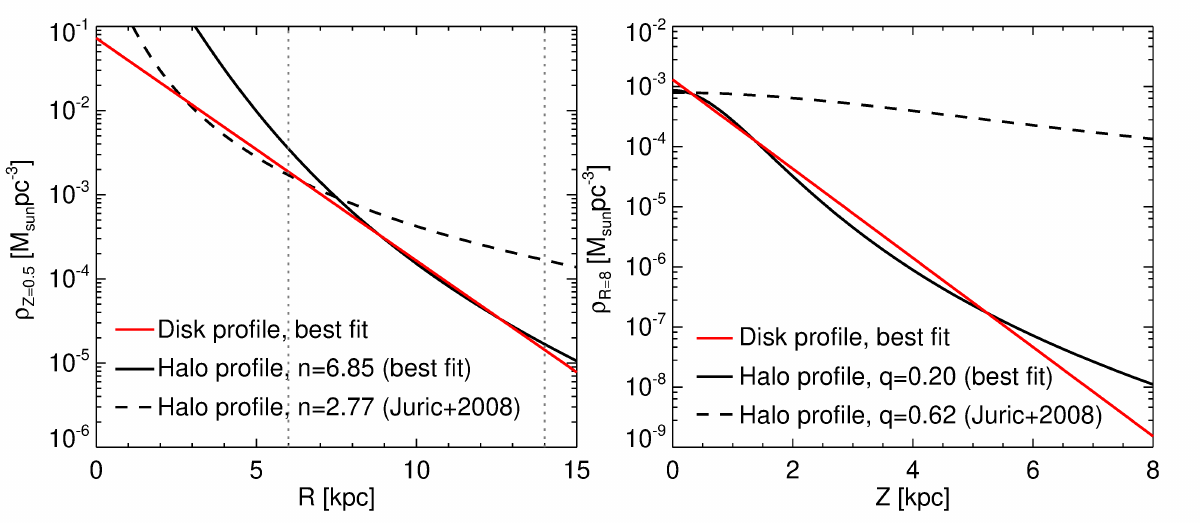}
    \caption*{{\bf Supplementary Figure 4 $|$ Stellar density distribution of different profiles for a mono-age stellar population of $\tau=10.5$~Gyr.} The left panel shows the radial variations of stellar density, while the right panel show the vertical variations of stellar density. The red and black solid lines are the best fits to the mono-age stellar population with disk and halo profile, respectively. The dashed line shows the canonical stellar halo profile \cite{Juric2008}, normalized by stellar mass density at the solar radius. The vertical dashed lines in the left panel delineates the radial regimes covered by our stellar sample. The left panel shows results at fixed height $Z=0.5$~kpc, while the right panel shows results at fixed radius $R=8$~kpc. The fits to the data using a halo yield comparable results to that of the disk profile in the radial range covered by the data, both different significantly to the canonical stellar halo. In the inner part ($R<6$), fitting the data with the halo profile however yields too large density.}
\end{figure}

\begin{figure}
    \center
    \includegraphics[width=1.0\textwidth]{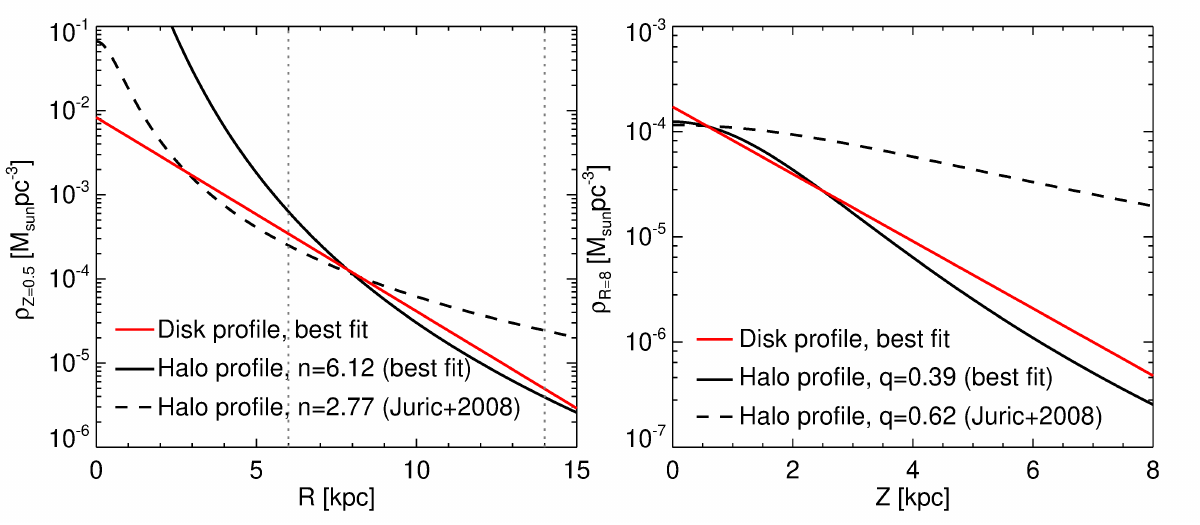}
    \caption*{{\bf Supplementary Figure 5 $|$ Stellar density distribution of different profiles for a mono-age stellar population of $\tau=14$~Gyr.} The left and right panels show the radial and vertical variations of stellar density, respectively. The red and black solid lines are the best fits to the mono-age stellar population with disk and halo profile, respectively. The dashed line shows the canonical stellar halo profile \cite{Juric2008}, normalized by stellar mass density at the solar radius. The vertical dashed lines in the left panel delineates the radial regimes covered by our stellar sample. The left panel shows results at fixed height $Z=0.5$~kpc, while the right panel shows results at fixed radius $R=8$~kpc. The fits to the data using a halo yield comparable results to that of the disk profile in the radial range covered by the data, both different significantly to the canonical stellar halo. In the inner part ($R<6$), fitting the data with the halo profile however yields too large density.}
\end{figure}

We have also examined a flared disk model  
\begin{align}
\rho(R,Z) = \rho_{R_\odot,0}\exp\left[-\alpha\cdot(R-R_\odot) \right] \exp \left[-\frac{\vert Z-Z_0\vert}{f(R)\times H_{Z}}\right],
\end{align}
\begin{align}
f(R) = \exp\left[\beta(R-R_\odot)\right],
\end{align}
where an exponential function $f(R)$ is adopted to describe the flaring effect, i.e., the increase of scale height as a function of $R$. The strength of the flaring effect is described by the parameter $\beta$.
The resultant structural parameters of such a flared disk model has been presented in the online article (\textbf{Extended Figure~5}).    

In the presence of disk flare, the $H_Z/H_R$ varies with Galactocentric radius. In order to have an intuitive description of the disk geometry, in \textbf{Supplementary Figure 6} we presented $H_Z$ and $H_R$ for three different radii, namely $R=2H_R$, $R=6$~kpc, and $R=R_\odot$(8.178~kpc). It shows that the scale height is smaller than the scale length for most of the sub-populations at the inner part of the disk. At $R=2H_R$, the scale height is less than half the scale length even for a large fraction of the oldest and most metal-poor sub-populations. At $R=6$~kpc, these old and metal-poor sub-populations still present a scale height smaller than scale length. At the $R=R_\odot$, the scale height is smaller than the scale length for sub-populations younger than $\sim11.5$~Gyr, but the scale height becomes comparable to the scale length for older sub-populations, and can even be larger than the scale length for the oldest sub-populations. As we expect that most stellar mass of the old, high-$\alpha$ disk populations are confined in a relatively small Galactic radius, we conclude that imposing the disk flare in the modelling does not impact the conclusions in the main text.

\begin{figure}
    \center
    \includegraphics[width=0.8\textwidth]{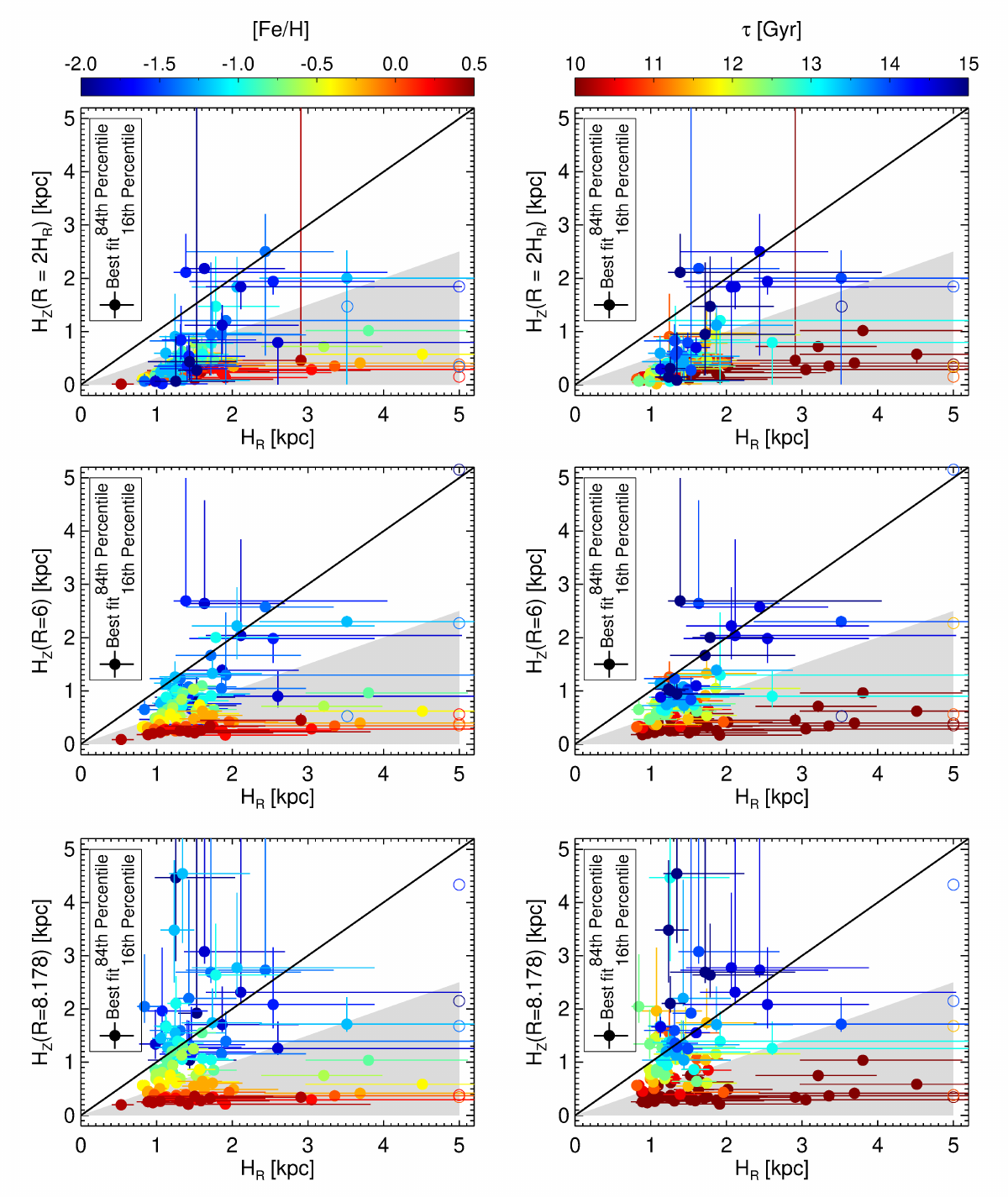}
    \caption*{{\bf Supplementary Figure 6 $|$ Scale height versus scale length of the high-$\alpha$ disk.} The scale height and length are derived by fitting the flared disk model. The left and right panels are color-coded by metalicity and age, respectively. The open circles without error bars in the figure refer to results where the MCMC fit failed to yield a valid parameter constraint. The shaded regions mark the parameter space where the scale height is smaller than half the scale length. Because of the existence of disk flare, the scale height varies with Galactocentric radius. From top to bottom, the results for three different radii, $R=2H_R$, $R=6$~kpc, and $R=R_\odot$(8.178~kpc) are shown. }
\end{figure}

\end{document}